\documentclass[aps,prl,showpacs,twocolumn,groupedaddress]{revtex4-1}
\usepackage{times,xspace}
\usepackage{graphicx,graphics,color,epsfig}
\usepackage{amsmath}
\usepackage{amssymb}
\usepackage{amsfonts}
\usepackage{amsfonts}
\usepackage{epstopdf}
\usepackage{bm}
\usepackage{hyperref}
\usepackage{color}
\usepackage{float}
\usepackage{braket}
\usepackage[normalem]{ulem}

\def\be{\begin{equation}}
\def\ee{\end{equation}}
\def\ba{\begin{eqnarray}}
\def\ea{\end{eqnarray}}

\newcommand{\TCC}{TlCuCl$_3$~}

\newcommand{\blue}{\textcolor{black}}

\begin{document}

\title{Higgs-Axion \blue{interplay} and anomalous magnetic phase diagram in TlCuCl$_3$}


\author{Gaurav Kumar Gupta$^1$}
\author{Kapildeb Dolui$^1$}
\author{Abhinav Kumar$^2$}
\author{D. D. Sarma$^2$}
\author{Tanmoy Das$^1$}
\affiliation{$^1$ Department of Physics, Indian Institute of Science, Bangalore, India - 560012}
\affiliation{$^2$SSCU, Indian Institute of Science, Bangalore, India - 560012}


\date{\today}

\begin{abstract}
What is so unique in TlCuCl$_3$ which drives so many unique magnetic features in this compound? To study these properties, here we \blue{employ} a combination of {\it ab-initio} band structure, tight-binding model, and an effective quantum field theory. Within a density-functional theory (DFT) calculation, we find an unexpected bulk Dirac cone without spin-orbit coupling (SOC). Tracing back to its origin, we identify, for the first time, the presence of a Su-Schrieffer-Heeger (SSH) like dimerized Cu chain lying in the \blue{3D} crystal structure. The SSH chain, combined with SOC, stipulates an anisotropic 3D Dirac cone where chiral and helical states are intertwined. As a Heisenberg interaction is introduced, we show that the dimerized Cu sublattices of the SSH chain condensate into spin-singlet, dimerized magnets. In the magnetic ground state, we also find a topological phase, distinguished by the axion \blue{angle}. Finally, to study how the topological axion term couples to magnetic excitations, we derive a Chern-Simons-Ginzburg-Landau action from the 3D SSH Hamiltonian. We find that axion term provides an additional mass term to the Higgs mode, and a lifetime to paramagnons, which are independent of the quantum critical physics. \blue{The axion-Higgs interplay can be probed with electric and magnetic field applied parallel or anti-parallel to each other.}
\end{abstract}


\maketitle
TlCuCl$_3$ has maintained a steady theme of research interests for more than two decades due to its unconventional magnetic properties. This material simultaneously accommodates several unusual magnetic properties, which are either individually present in other magnetic systems, or even absent. \TCC  is paramagnetic at ambient condition, but undergoes a quantum phase transition to an antiferromagnetic (AFM) state with small pressure\cite{pressure4,pressure3,pressure1,pressure2}, or with magnetic field\cite{mag4,mag3,bec2,mag2} or with nonmagnetic impurity\cite{impurity}. (a) The AFM phase of TlCuCl$_3$ arises from the formation of nearest neighbor quantum dimer, a spin-singlet excitation often seen in spin-liquid systems, and it does not necessarily break translational symmetry.\cite{th21,bec1,th1,pressure1,mag1} (b) Higgs mode was postulated to be associated with a larger class of continuous symmetry breaking order parameters\cite{th21,Higgs}, but rarely observed due to its evanescent characteristics.\cite{Higgsvisibility1,Higgsvisibility2,Higgsvisibility3} TlCuCl$_3$ is one of the earlier systems where a Higgs mode was observed in the AFM phase, in addition to one gapless and one gapped Goldstone modes.\cite{pressure1,pressure2,pm1} (c) Paramagnons, gapped magnetic excitations in non-magnetic phase, usually have short lifetime, as they decay into the particle-hole continuum. But in TlCuCl$_3$, paramagnons have equally large lifetime as that of the Higgs mode across the critical point.\cite{pressure2,pm1} (d) In this material, Bose-Einstein condensation of spin-excitations was experimentally achieved.\cite{bec1,bec2,mag1,th1} Therefore, TlCuCl$_3$ provides an important playground to drive such a wide variety of unusual magnetic properties within the same crystal. 

Considerable experimental and theoretical studies have been devoted to understand these \blue{unusual} magnetic properties of \TCC \cite{bec1,bec2,pm1,pressure1,pressure2,mag1,mag2,th1,th2,th21,th3,th4,Sushkov,QMC}. In various theoretical models, the Heisenberg type spin-spin interaction is mainly considered\cite{th1,th2}, consistently explaining the formation of spin-singlet dimers, and reproducing the experimental spin-wave dispersion\cite{pressure1,pressure2,mag1}. Within the so-called $\phi^4$-theory, one can also obtain a characteristic scale of the Higgs mode's lifetime\cite{th1,th2,th21,Higgslifetime,Higgsvisibility3}. 

To look into these questions from a materials specific, microscopic perspective, we investigate the magnetic properties of \TCC~constrained by its DFT band structure. To our surprise, we find that there exists an isolated Dirac cone in the bulk band structure, even in the absence of spin-orbit coupling (SOC) and magnetism. The origin of such a Dirac cone is traced back to the presence of a Cu-chain along the \blue{$c$}-direction, which is reminiscent of the celebrated Su-Schrieffer-Heeger (SSH) chain, so far known to exist in 1D polyacetylene chain.\cite{SSH} The SSH chain can produce an 1D Dirac-like degenerate point at $k_z=\pm\pi/\blue{c}$. However, the DFT result shows a single band crossing point at ${\bf k}=(0,0,\pm\pi/\blue{c})$. We develop a 3D SSH model for this system, which reproduces the anisotropic 3D Dirac cone with chiral (sublattice-momentum locking) states along the $k_z$-direction and helical (spin-momentum locking) state in the basal plane.

As the AFM order turns on, we find that the spin-singlet dimers are formed between the nearest neighbor Cu-sublattices of the SSH chain. This causes an inversion of the helicity between the two Cu-sublattices, driving a topologically non-trivial phase, as distinguished by a finite axion angle ($\theta$) within the Chern-Simon theory. \blue{The axion term introduces a positive/negative magneto-electric effect, which couples the parallel/antiparallel components of the electric and magnetic fields.} The interplay between the topological excitations (axions) and magnetic excitations (mainly Higgs, paramagnons modes) is studied here within a microscopically derived Chern-Simons-Ginzburg-Landau (CSGL) model. \blue{We find that (a) the axion term adapts a second order phase transition to a first-oder one for positive/negative magneto-electric coupling, respectively. (b) The axion term gives a new contribution to the Higgs mass and lifetime terms which are independent of the magnetic order parameter, and hence can gap out the Higgs mode at the AFM critical point. (c) The N\'eel temperature is increased (decreased) with positive (negative) magneto-electric coupling. Such a topological nature of the AFM phase can be verified by applying electric field parallel to the magnetization or applied magnetic field.}

\begin{figure}[t]
	\includegraphics[width=70mm]{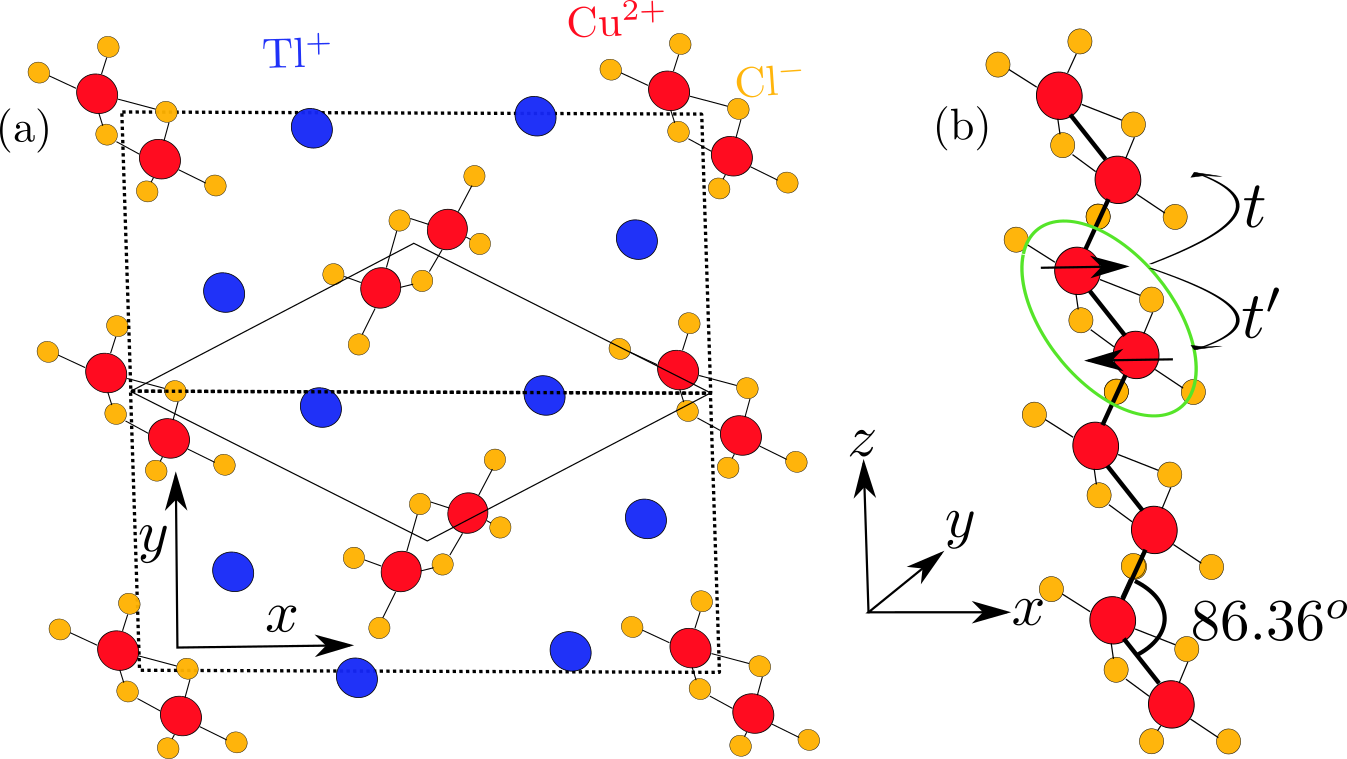}
	\caption{Top (a) and side (b) views of TlCuCl$_3$ unit cell. Each unit cell contains two SSH chains of Cu-atoms (red symbols), which are mutually rotated by 90$^o$.  The smaller rectangle connecting nearest-neighbor SSh chain gives the conventional unit cell we invoke to obtain a 2-band model for the Dirac cone. (b) A single SSH chain is shown with \blue{$t,t^\prime$ representing inter-sublattice hoppings along the $\pm z$ directions, respectively. Arrows dictate in-plane spin-polarization in the two Cu sublattices, forming a singlet dimer in the AFM phase}.} 
\label{fig:structure}
\end{figure}

{\it DFT results.} TlCuCl$_3$ crystallizes in the monoclinic P2$_1$/c space group, with 4 formula units per unit cell. We use the experimental lattice constants of $a$ = 14.144 \AA, $b$= 8.89 \AA, and $c$= 3.983 \AA , and $\alpha$ = 96.32$^\circ$. The top view in Fig.~\ref{fig:structure}(a) shows a rectangular projection of the unit cell on the \blue{$xy$}- plane. Each formula unit contains two inequivalent SSH chains along the \blue{$z$}-axis as shown in Fig.~\ref{fig:structure}(b), at the center and corners of the rectangle. Because of different Cl-environments, the two nearest neighbor distance between Cu-Cu atoms become slightly different resulting in a SSH structure. 

\begin{figure}[h]
	\includegraphics[scale=.45]{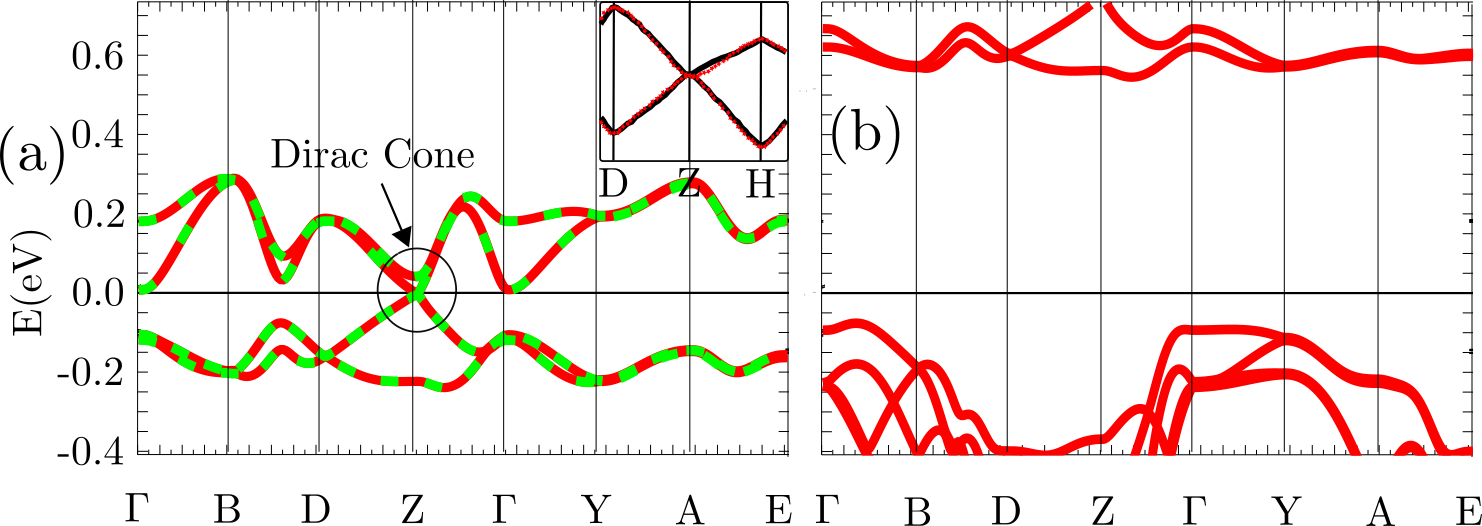}
	\caption{(a) DFT band structure of TlCuCl$_3$, plotted along high-symmetric directions.\cite{footHSDir}. Green dashed and red solid lines depict the bands calculated without and with SOC, respectively. Since SOC is of the order of ~5 meV, the band splitting is not visible in this energy scale. {\it Inset:} Fittings of the 3D SSH model near the Dirac cone. Red line represents DFT bands while black line gives TB fitted bands. An extra point, H(0,$\pi$,0), is used for fitting to capture true 3D nature of the dirac cone. (b) \blue{DFT band structure in the AFM phase}.}
\label{fig:DFT}
\end{figure}

We compute the DFT band structure using the Local Density Approximation (LDA) exchange correlation as implemented in the Vienna  ab-initio simulation  package  (VASP)\cite{DFT1,DFT2}. LDA+U ($U=4$ eV) method is used to deal with the strong correlation features on Cu-3d orbitals. The non-magnetic DFT band structure in Fig.~\ref{fig:DFT} shows four bands near the Fermi level ($E_F$), stemming from the $d$-orbitals of the Cu-atoms. Each SSH chain is individually responsible for forming a 1D Dirac cone at the $k_z=\pm\pi$-point. The inter-chain hopping breaks the degeneracy of the bands, resulting in two gapped bands, and one single Dirac cone. \blue{The Dirac cone is also obtained in an earlier LMTO-based DFT calculation\cite{LMTODFT}, and is also robust to GGA functional (not shown), and is reproducible with different values of $U$ (see SM\cite{SM}).}

{\it Tight-binding model.} Our main interest is to study the topological properties arising from the bulk Dirac cone. Since there is only a single Dirac cone present near $E_F$, the minimal model required to capture the essential topological properties is a two-band model forming the Dirac cone. We therefore start with a two-band tight-binding model, coming from the Cu-sublattices in a given SSH chain, and allow inter-chain hoppings in all three dimensions. We may refer the corresponding model as a 3D SSH model. 

In what follows, we work in a single Cu-chain per conventional unit cell, as indicated in ~Fig.~\ref{fig:structure}(a). We express the corresponding Hamiltonian in a 2-components spinor as $\Psi({\bf k}) = (\psi_{\rm A}({\bf k}), \psi_{\rm B}({\bf k}))^{T}$, where `A', and `B' stand for two Cu-atoms as
\begin{eqnarray}
H_{0}=\sum_{{\bf k},{ij}\in ({\rm A,B})}\xi^{ij}_{\bf k}\psi^{\dag}_{i}({\bf k})\psi_{j}({\bf k}).
\label{Eq:H0}
\end{eqnarray}
Here $\xi^{\rm AA}_{\bf k}=\xi^{\rm BB}_{\bf k}$, and $\xi^{\rm AB}_{\bf k}$ are the intra-, and inter-sublattice dispersions, respectively. 
The energy eigenvalues are $E_{\bf k}^{\pm}=\xi^{\rm AA}_{\bf k} \pm |\xi^{\rm AB}_{\bf k}|$. The two bands meet at the locii of $|\xi^{\rm AB}_{\bf k}|=0$, while $\xi^{\rm AA}_{\bf k}$ gives an overall shift of the degenerate points in energy.  

In the case of an isolated 1D SSH chain, $\xi^{\rm AB}_{k_z}$ is often described by $\xi^{\rm AB}_{k_z}\rightarrow \left(t+t'e^{-ik_z}\right)$, where $t$, and $t'$ are the inter-sublattice hoppings along the $\pm z$-direction, respectively [see Fig.~\ref{fig:structure}(b)]. A Dirac cone forms at $k_z=\pm\pi$ when {\color{blue}$t'= t$}. In the same spirit, we cast the Hamiltonian in Eq.~\ref{Eq:H0} into a 3D SSH model as 
\begin{eqnarray}
\xi^{\rm AB}_{\bf k}&=&T_{{\bf k}_{\perp}}+T'_{{\bf k}_{\perp}}e^{-ik_z},
\label{xiAB}
\end{eqnarray}
where ${\bf k}_{\perp}=(k_x,k_y)$. $T'_{{\bf k}_{\perp}}$, and $T_{{\bf k}_{\perp}}$ have the same meanings as $t'$, and $t$, but due to inter-SSH chain hoppings, they acquire in-plane dispersions. $\xi^{\rm AA}_{\bf k}$,  $T_{{\bf k}_{\perp}}$, and $T'_{{\bf k}_{\perp}}$ are expressed in terms of the Slater-Koster tight-binding (TB) hopping integrals between intra-, and inter-chain hoppings, and we spare the details to SM\cite{SM}.

Following the DFT result, we fit the TB dispersions to the DFT band with the constraint that $T^\prime_{{\bf k}^*}=T^{}_{{\bf k}^*}$ only at ${\bf k}^*=(0,0,\pi)$. Hence we reproduce a single band crossing, with linear dispersion in $\delta k_z$, and quadratic dispersion in ($\delta k_x$, $\delta k_y$), where $\delta{\bf k}={\bf k}^*-{\bf k}$ and $\delta{\bf k}<<1$ (see {\it inset} to Fig.~\ref{fig:DFT}). 


{\it SOC.} Although SOC is weak here, it is however sufficient to introduce helicity in the low energy spectrum. In \TCC, spins are aligned in the ${\bf k}_{\perp}$-plane near the critical point, consistent with experiments\cite{pm1,pressure1,pressure2,mag1} and DFT calculation, see below. This also makes the in-plane SOC to be dominant. A full derivation of the SOC is given in the SM\cite{SM}, and its non-vanishing component is given by
\begin{eqnarray}
H_{\rm SOC}&=&\sum_{i,j\in (A,B)}\sum_{{\bf k},ss'}\left[\psi^{\dag}_{i,s}({\bf k})\left({\bm \alpha}^{ij}_{\bf k}\times {\bm \sigma}_{ss'}\right)\psi_{j,s'}({\bf k}) \right].
\label{Eq:HSOC}
\end{eqnarray}
$s,s'$ give spin components. The components of the velocity operators are ${\bm \alpha}^{ij}_{\bf k} =\alpha^{ij}_0\left(-\partial_{k_y} \xi^{ij}_{\bf k},\partial_{k_x} \xi^{ij}_{\bf k},0\right)$, with $\alpha^{ij}_0$ being the corresponding SOC strengths. Eq.~\ref{Eq:HSOC} allows several SOC terms, however, fitting to DFT results indicate that $\alpha^{\rm AB}_0\rightarrow 0$, and $\alpha^{\rm AA}_0=0.05$ eV.\blue{\cite{foot_TBfit}}

{\it Dirac Hamiltonian:} To proceed further, it is convenient to express the Hamiltonian [Eqs.~\ref{Eq:H0}, and \ref{Eq:HSOC}], in the Dirac matrix form. We take the spinor $\Psi({\bf k}) = (\psi_{{\rm A}\uparrow}, \psi_{{\rm B}\uparrow}, \psi_{{\rm A}\downarrow}, \psi_{{\rm B}\downarrow})^{T}$ to obtain
\begin{eqnarray}
H({\bf k})=\xi^{\rm AA}_{\bf k}\mathbf{1}_{4\times 4}+ \sum_{i=1}^{5} d_i({\bf k})\Gamma_i.
\label{ham}
\end{eqnarray}
where $\Gamma=(\sigma_x\otimes\mathbf{1},\sigma_y\otimes\mathbf{1},\mathbf{1}_{2\times 2}\otimes\tau_x,\mathbf{1}_{2\times 2}\otimes\tau_y,\sigma_z\otimes\tau_z)$, where $\sigma$, and $\tau$ are the Pauli matrices in the spin and sublattice basis, respectively. The components of the $d$-vectors are ${\bf d}$ = ($\alpha'^{\rm AA}_x$,$-\alpha''^{\rm AA}_y$,$\xi'^{\rm AB}_k$, $\xi''^{\rm AB}_k$,0). $H({\bf k})$ is invariant under both time-reversal and parity symmetries.\cite{footparity}

{\it AFM calculations.} Next, we perform spin-polarized DFT calculations with and without the SOC within the LSDA (local spin-density approximation) method by using VASP package. \blue{The spin-configuration is taken to be non-collinear.} We find a AFM ground state with antiparallel spin between `A' and `B' sublattices of the SSH chain\cite{foot_pressure}. We find that the spins are quantized in the $xy$ plane, as seen in experiment\cite{pm1,pressure1,pressure2,mag1}, and the easy axis is almost along the diagonal direction in this plane. The DFT predicted magnetic moment along the $z$-direction is negligibly small, and that in the $ab$-plane are $m_{\rm A,B}=\pm 0.43\mu_{\rm B}$ for the two Cu atoms, respectively. The magnetic band structure shows insulating behavior with a band gap $\sim$ 1.3 eV.\cite{SM,foot_surface} From the band gap and magnetic moment, we estimate the AFM coupling to be around $J\sim$ 1.5~eV, which is close to the value estimated in neutron scattering measurement.\cite{ExpINSJ,ExpINSJ_2}

\blue{Since the magnetic moment is small, we take an itinerant model of the AFM phase. Guided by the DFT result of an easy-axis quantization of the spins in the AFM phase, we specialize the Heisenberg interaction only along the spin-quantization axis $S^{z}$, and between the nearest `A' and `B' sublattices only:} $H_{\rm I}=J\sum_{\Braket{i,j}\in{\rm (A,B)}}S^{z}_i S^{z}_j$, where $S^{z}_i$ is the spin operator. 
The AFM order parameter is defined as $\phi=(m_{\rm A}-m_{\rm B})/2$, where the magnetization is $m_{\rm A/B}=\langle{S^{z}_{\rm A/B}}\rangle$. The excitation energy gap in the band structure is $\Delta=J\phi$. Such an order parameter has been used earlier in \TCC, and is found via self-consistent calculation to define the AFM ground state.\cite{QMC} Using Hubbard-Stratonovic decomposition of the $H_{\rm I}$, we obtain the magnetic perturbation\cite{SM}
\begin{eqnarray}
H_{\rm I}&\approx &J\phi\Gamma_5=d_5\Gamma_5.
\label{Eq:HI}
\end{eqnarray}

\begin{figure}
\includegraphics[scale=.47]{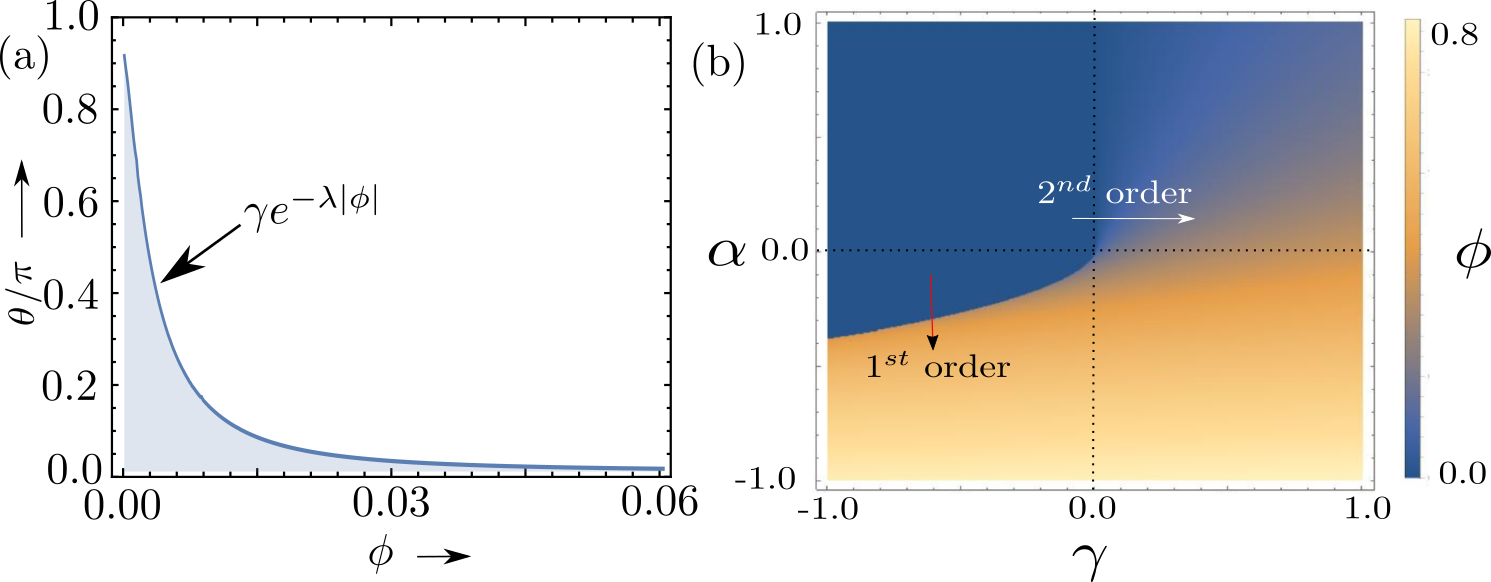}
\caption{(a) Computed values of the axion angle $\theta$ as a function of the magnetic order parameter $\phi$. (b) Color plot depicts values of magnetization as a function of the GL coefficient $\alpha$, and the CS coefficient $\gamma$ from Eq. 9. We set $\beta>0$. Horizontal arrow (white) indicates a second order phase transition line where order parameter decreases continuously, while vertical arrow dictates a first-order phase transition line.}
\label{axion_angle}
\end{figure}

{\it Helicity inversion and topological axion insulator.} The AFM order introduces a crucial change in the SOC term in Eq.~\eqref{Eq:HSOC}. Since the spin polarization is reversed between the `A' and `B' sublattices, the corresponding SOC is also reversed, i.e., $\alpha^{\rm AA}_{\bf k}=-\alpha^{\rm BB}_{\bf k}$. This induces an inversion in the helicity between the `A' and `B' sublattices. This helicity inversion endows the system to acquire a non-trivial topological phase.\cite{TDNC,SODW,TDRMP} We incorporate the helicity inversion by changing $\Gamma_{1,2}\rightarrow\sigma_x\otimes\tau_z,\sigma_y\otimes\tau_z$. 

The topological invariant of a 3D AFM insulator cannot be defined by the usual $\mathbb{Z}_2$ invariant or Chern number, but by \blue{a magneto-electric coupling} with the coupling constant proportional to the `axion \blue{angle}' $\theta$\cite{zhang_prb,JMoore,axionJPSJ}. The axion \blue{angle} ($\theta$) is the $\mathbb{Z}_2$ invariant (multiplied by $\pi$) for a time-reversal invariant system, and vanishes continuously as the magnetization increases\cite{JMoore,zhang_nat}. The axion \blue{angle} is the solid angle enclosed in the $d$-space as one encircles the entire 3D Brillouin zone.\cite{zhang_nat,axionJPSJ} Reminiscence to the topological phase transition in a single SSH chain, we also find here that $\theta$ becomes finite when the zeros of $d_3({\bf k})={\xi^{\prime{\rm AB}}_{\bf k}}$ lies inside the solid angle, giving the condition that $T_{{\bf k}_{\perp}}\le T'_{{\bf k}_{\perp}}$, for ${\bf k}\in {\rm BZ}$. Having a Dirac cone in the SOC band structure, we ensure that such a condition is automatically satisfied in the non-interacting phase. 

For $\phi\rightarrow 0$, we obtain $\theta=\pi$ (see SM\cite{SM} for the axion calculation details). For finite $|\phi|>0$, we numerically find that $\theta$ decreases exponentially as shown in Fig.~\ref{axion_angle}a, as
\begin{equation}
\theta=\pi e^{-\lambda |\phi|}, 
\label{axion}
\end{equation}
where $\lambda\propto J$, is a fitting parameter, obtained to be $\lambda=$220. \blue{Large value of $\lambda$ indicates that $\theta$ decreases very rapidly with $\phi$}. 
Owing to time-reversal symmetry breaking, the corresponding topological axion phase does not exhibit any gapless edge state.\cite{foot_surface}

{\it Chern-Simons-Ginzburg-Landau analysis:} Finally we discuss the implications of the topological excitations to the magnetic properties. The topology induced axion excitations are described by a Chern-Simons (CS) term in the effective Lagrangian.\cite{zhang_prb,JMoore} On the other hand, the interaction induced  magnetic excitations are captured within the Ginzburg-Landau (GL) theory. The field-theory description of the competition between electronic interaction and topological responses due to probe electromagnetic fields $(A_0, {\bf A})$ is developed earlier in the context of fractional quantum Hall effect,\cite{CSGL} and is termed as Chern-Simons-Ginzburg-Landau (CSGL) theory. In addition to probe fields, there may arise intrinsic `statistical' gauge fields $(a_0, {\bf a})$. Thanks to the linear combination form of the intrinsic and probe gauge fields in the Lagrangian, we can combine their effects in a total gauge field as $\mathcal{A}_0=a_0+A_0$, and $\bm{\mathcal{A}}={\bf a}+{\bf A}$. The full Lagrangian density can be split into four parts\cite{CSGL,foot_CSGL} $\mathcal{L}_{\rm total}=\mathcal{L}_{\rm KE} + \mathcal{L}_{\rm MW} + \mathcal{L}_{\rm GL} + \mathcal{L}_{\rm CS}$. $\mathcal{L}_{\rm KE}$ is the kinetic energy and $\mathcal{L}_{\rm MW}$ is the Maxwell term. Since these two terms do not contribute to the magnetic phase diagram and Higgs mode, we do not include them henceforth.\cite{SM} The remaining GL and CS terms can be derived using the path integral description of coherent states of the total Hamiltonian $H_0+H_{\rm SOC}+H_{\rm I}$, and then integrating out the fermionic degrees of freedom (see SM\cite{SM}) to obtain
\begin{eqnarray}
%
\mathcal{L}_{\rm GL} = \blue{-}\alpha|\phi|^2 \blue{-} \beta|\phi|^4, ~~~\mathcal{L}_{\rm CS} = \theta \frac{\hbar}{\Phi_0^2}{\bf E}\cdot{\bf B},
\label{Eq:CS}
\end{eqnarray}
Here $\alpha$, and $\beta$ are the GL-coefficients, arise from the spin-susceptibilities \blue{and depend on the band structure parameters and SOC}, as explicitly evaluated in SM\cite{SM}. $\Phi_0=h/e$ is the magnetic flux quanta. ${\bf E}$ and ${\bf B}$ are electric and magnetic fields corresponding to $\bm{\mathcal{A}}$. 

Apparently, there is no direct coupling between the scalar field $\phi$ and the axion mode $\theta$, rather the axion field $\theta$ directly stems from the scalar field $\phi$, Eq.~\eqref{axion}. Substituting for $\theta$ in Eq.~\ref{Eq:CS}, we get $\mathcal{L}_{\rm CS}=\gamma e^{-\lambda|\phi|}$, where \blue{$\gamma=-\frac{\pi\hbar}{\Phi_0^2}{\bf E}\cdot{\bf B}$ is a variational parameter. $\gamma>0$ ($\gamma<0$) if ${\bf E}$ and ${\bf B}$ are parallel (antiparallel) to each other, and otherwise zero.} Neglecting the irrelevant space-time dependence of the order parameter, we arrive at the CSGL term, expressed exclusively in terms of the AFM field $\phi$ as
\begin{eqnarray}
\mathcal{L}_{\rm CSGL}&=&-\alpha|\phi|^2 - \beta|\phi|^4 - \gamma e^{-\lambda|\phi|} + \gamma.
\label{CSGL}
\end{eqnarray}
(We have added a constant term $\gamma$ to shift the Free energy ($\propto-\mathcal{L}$) minimum to zero at $\phi=0$). The magnetic phase transition, and magnetic excitations can now be studied as a function of four variational parameters $\alpha, \beta, \gamma$, and $\lambda$. 

{\it Magnetic phase diagram:} Minimization of $\mathcal{L}_{\rm CSGL}$ occurs at a finite value of $\phi=\phi_0$, which are the root of the following secular equation:
\begin{equation}
2\left(\alpha+2\beta|\phi_0|^2\right)|\phi_0| = \gamma\lambda e^{-\lambda|\phi_0|} .
\label{Eq:Lminimum}
\end{equation}
Solution of the above equation is non-trivial to manage analytically. For $\gamma\rightarrow 0$, we recover the typical GL result of $|\phi_0|=\sqrt{-\alpha/2\beta}$, giving a second order phase transition as $\alpha$ becomes negative (with $\beta>0$). Since we are in the vicinity of a second order phase transition, we set $\beta>0$, and $\lambda= 220$ (from Fig.~3a). We study the solution of $\phi_0$ as a function of $\alpha$ and $\gamma$, as given in Fig.~3b.  For $\gamma>0$ region, we find that $\phi_0$ decreases {\it continuously} to zero, suggesting a second order phase transition as a function of both $\alpha$, and $\lambda$. On the other hand, for $\gamma<0$, we notice that the phase boundary from finite $\phi_0$ to zero is {\it discontinuous}, implying that the phase transition becomes first order. To understand this behavior, we expand the CS term in the leading order in $|\phi|$ as $-\gamma\lambda|\phi|$. So, for $\gamma>0$, $\mathcal{L}_{\rm CSGL}$ decreases with increasing $|\phi|$, and hence its minima continuously move from $\phi=0$ to $|\phi_0|>0$ $-$ a second order phase transition. While for $\gamma<0$, $\mathcal{L}_{\rm CSGL}$ increases with increasing $|\phi|$, and then a second minimum occurs at a finite $|\phi_0|>0$. Since the finite$|\phi_0|$ minima are now disjointed from the $\phi=0$ minimum, we have a first order phase transition. 

In both cases, we also observe that the phase boundary shifts from the GL limit of $\alpha=0$ line to finite values of $\pm\alpha$ in the two cases, respectively. This has implications to the values of the N\'eel temperature, and the Higgs mass. By expanding the axion term up to $|\phi|^2$, and assuming $\alpha=\alpha_0(1-T/T_{{\rm N},0})$ for $\gamma=0$, we obtain that the effective N\'eel temperature modifies as 
\begin{equation}
T_{\rm N}=T_{{\rm N},0}(1+\gamma\lambda^2/2\alpha_0).
\label{TN}
\end{equation}
$T_{\rm N}$ increases (decreases) for $\gamma>0$ ($\gamma<0$).  This means, $T_{\rm N}$ increases (decreases) as the \blue{applied} magnetic and electric fields are parallel (antiparallel), which can be used to verify the topological nature of this magnetic ground state. 

{\it Magnetic excitations.} Finally we study the interplay between the magnetic and topological excitations. We expand the order parameter near its expectation value as $\phi=|\phi_0 + \delta \phi(x)|e^{i\eta(x)}$, where $\delta \phi$, and $\eta$ are the corresponding amplitude, and phase fluctuations, respectively. In Eq.~\eqref{CSGL}, we find that $\mathcal{L}_{\rm CSGL}$ depends on the amplitude $|\phi|$ only, and thus the phase ($\eta$) fluctuations remain gapless (Goldstone modes) even in the presence of the axion term (see SM\cite{SM} for the derivation). [In fact, all Goldstone modes can be gauged out by a suitable gauge transformation of the EM fields $\mathcal{A}$.] Substituting $\phi=|\phi_0 + \delta \phi(x)|$ in Eq.~\eqref{Eq:CS}, we can estimate the mass of the amplitude mode as $M=\frac{1}{2}\partial^2_{\delta \phi}\mathcal{L}|_{{\delta \phi}=0}$. After substituting Eq.~\eqref{Eq:Lminimum} at the saddle point of the Lagrangian, we obtain the Higgs mass 
as
\begin{eqnarray}
M&=&2\alpha+12\beta|\phi_0|^2+\gamma\lambda^2e^{-\lambda |\phi_0|}.
\end{eqnarray}
For $\gamma\rightarrow 0$, we recover the GL value of $M=4\beta|\phi_0|^2$ vanishing at the critical point where $\phi_0\rightarrow 0$. However, in the present case, we find that there is a finite Higgs mass even above the critical point and eventually vanishes only when $\alpha=0$. On the other hand, for $\gamma>0$, we notice in Fig. 3b, that a continuous phase transition can occur at $\alpha>0$, giving a non-vanishing Higgs mass at the critical point, which may be called `topological paramagnons'. For $\gamma<0$, we have a first order phase transition at $\alpha<0$, where the order parameter is discontinuous, and thus also the Higgs mass must vanish discontinuously.  

Calculation of Higgs mode's lifetime is rather cumbersome. One source of Higgs lifetime is the quartic term in the Lagrangian. In this spirit, the leading term in the inverse lifetime ($\tau$) is proportional to the coefficient of the $\delta\phi^4$, which can be obtained from $\partial^4 \mathcal{L}_{\rm CSGL}/\partial \delta \phi^4|_{\delta \phi=0}$ term, leading to
{
\begin{eqnarray}
\frac{1}{\tau}&\propto& 24 \beta|\phi_0|-\lambda^3\gamma e^{-\lambda|\phi_0|}.
\label{lifetime}
\end{eqnarray}
}
Eq.~\eqref{lifetime} suggests that Higgs lifetime rather decreases near the critical point for $\gamma\ne 0$, while away from the critical point, as the second term becomes dominant, it tends to increase. Hence we can argue that the `topological paramagnons' have much reduced decay rate.

{\it Conclusions and outlook:} Since there is only one Higgs mode in this model, the axion-Higgs  coupling can be captured well within the proposed CSGL theorem, and the corresponding Lagrangian resemblance that of the Standard Model of the particle physics. It is known that in the case of a Higgs doublet, there arises axion-Higgs cross term in the Lagrangian, and the system looses its CSGL symmetry, and one obtains a so-called Peccei-Quinn (PQ) symmetry, which violates the Standard model.\cite{Axiondoublet} However, the predicted Higgs doublet is yet to be observed. Based on the above analysis, we anticipate that our work will stimulate research for the realization of PQ symmetry in condensed matter systems where topological axion and Higgs terms are intertwined.\cite{Axiondoublet}

\begin{acknowledgments} 
We thank Soumi Ghosh for help with some of the analysis. TD acknowledges the financial support from Science and Engineering Research Board (SERB), Department of Science \& Technology (DST), Govt. of India for the Start Up Research Grant (Young Scientist).
\end{acknowledgments}


\begin{thebibliography}{1}
\bibitem{foot_TBfit}We performed TB fitting to the DFT band structure in two steps. First, we fit the band structure without SOC, and obtained hopping parameters with gives a band degenerate [point at the Z-point. With SOC, we re-fit the entire band structure. We find that the hopping parameters remain essentially unaffected with SOC as the DFT band with SOC is also very much the same.
\bibitem{foot_pressure}In the DFT calculation, we used experimental lattice constant which produces a  small pressure 0.5 GPa in the unit cell, and thus it mimics the experimental condition of pressure tunned AFM transition. This is also done in Ref.~\cite{LMTODFT}. \blue{The band structure with relaxed coordinate shows an inverted band structure near the Z-point, signifying that the system becomes topologically non-trivial at zero pressure}.
\bibitem{foot_surface} The magnetic state represents a non-trivial topological axion insulator in the small magnetic moment region near the critical point. However, owing to the loss of time-reversal symmetry, the surface state does not host any gapless state. We do not expect to observe any edge state in the DFT calculation.
\bibitem{foot_CSGL}Additional CS terms related to anyons,\cite{CSGL} and the Maxwell terms are not included since they do not directly impact the field $\phi$ in the lowest orders.
\end{thebibliography}
\end{document}


\relax

\title{\titlename}
\title{Supplementary Material: Higgs-Axion conversion and anomalous magnetic phase diagram in TlCuCl$_3$}

\author{Gaurav Kumar Gupta$^1$}
\author{Kapildeb Dolui$^1$}
\author{Abhinav Kumar$^2$}
\author{D. D. Sarma$^2$}
\author{Tanmoy Das$^1$}
\affiliation{$^1$ Department of Physics, Indian Institute of Science, Bangalore, India - 560012}
\affiliation{$^2$ Solid State and Structural Chemistry Unit, Indian Institute of Science, Bangalore, India - 560012}

\date{\today}

\maketitle
In this supplementary material, we give details of the derivations corresponding to various terms presented in the main text.  In Sec. I, we show DFT calculations including spin-orbit coupling in the paramagnetic and AFM phases. Next we derive the Ginzburg-Landau-Chern-Simons theory. Subsequently, we calculate the Lagrangian minima and AFM transition temperature  for the CSGL theory. In the last section, calculations of Higgs mass and corresponding lifetime are presented.

\section{DFT calculation W/ SOC}
We compute the DFT band structure using the Local  Density Approximation (LDA) exchange correlation as implemented in the Vienna  ab-initio  simulation  package  (VASP)\cite{DFT1}. The results remain characteristically the same for the GGA and other functionals. The DFT band structure also agrees well with a previous LMTO calculation.\cite{LMTODFT} In our LDA+U calculation, the electronic wave function is expanded using plane wave up to a cutoff energy of 500 eV. Brillouin zone sampling is done by using a ($8\times8\times8$) Monkhorst-Pack k-grid. Projected augmented-wave (PAW) pseudo-potentials are used to describe the core electron in the calculation\cite{DFT2}.

\begin{table}[H]
\caption{Table to show the experimantal and relaxed lattice parameters}
\begin{center}
	\begin{tabular}{|c|c|c|c|c|c|c|}
		\hline
		& a (\AA) & b (\AA) & c(\AA) & $\alpha$($^o$) & $\beta$($^o$)&$\gamma$($^o$) \\
		\hline
		Experimental & 14.144 & 8.890 & 3.982 & 83.68 & 90  & 90\\
		\hline
		Relaxed &13.587 &8.651 &3.886 &84.261 &89.983 &90.01 \\
		\hline
	\end{tabular}
\end{center}
\end{table}
We found in the DFT band structure that the magnetic gap is larger than the crystal field splitting (CFS), and thus one obtains a band insulating behavior in the electronic properties. As one approaches the magnetic critical point, the magnetic gap takes over for $\Delta<\Delta_{\rm CFS}$.

\begin{figure}
	\begin{center}
	\includegraphics[scale=0.7]{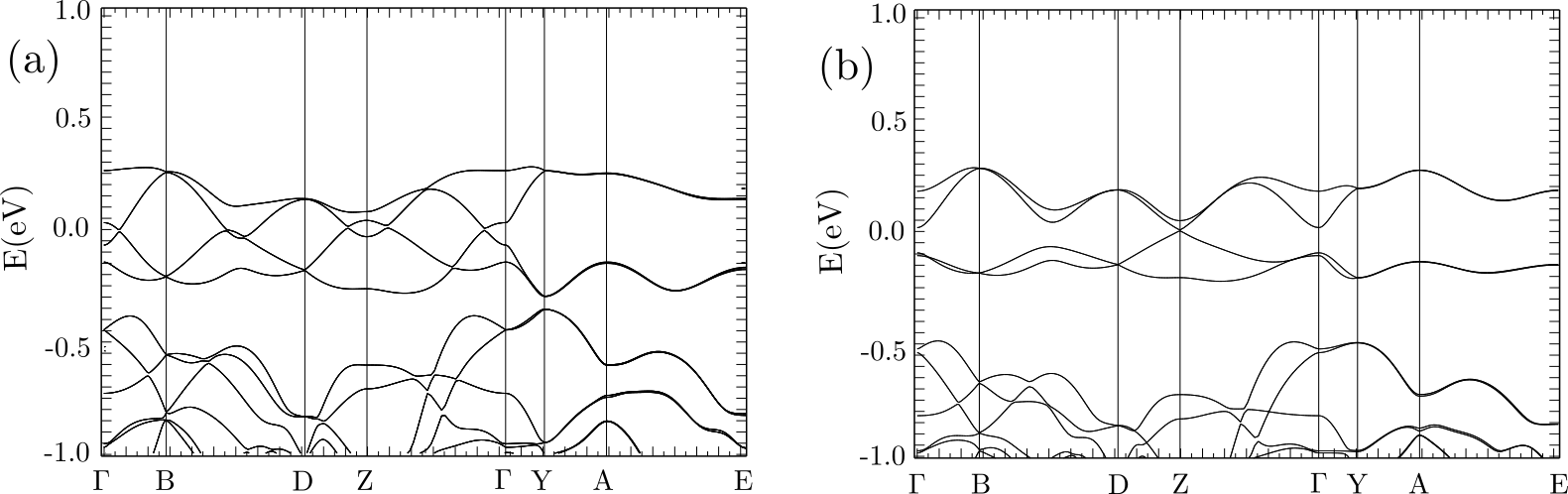}
	\caption{Comparison of band structure with relaxed (a) and experimental (b) lattices constant. In the relaxed parameter case, we find that the bands are inverted at Z-point ($t^\prime<t$), while it gives a Dirac cone ($t^\prime\sim t$). Therefore, we anticipate the topological nature should be persistent to ambient pressure, however the magnetic order disappears here.}
		\end{center}
\end{figure}

\begin{figure}[h]
	\includegraphics[scale=0.7]{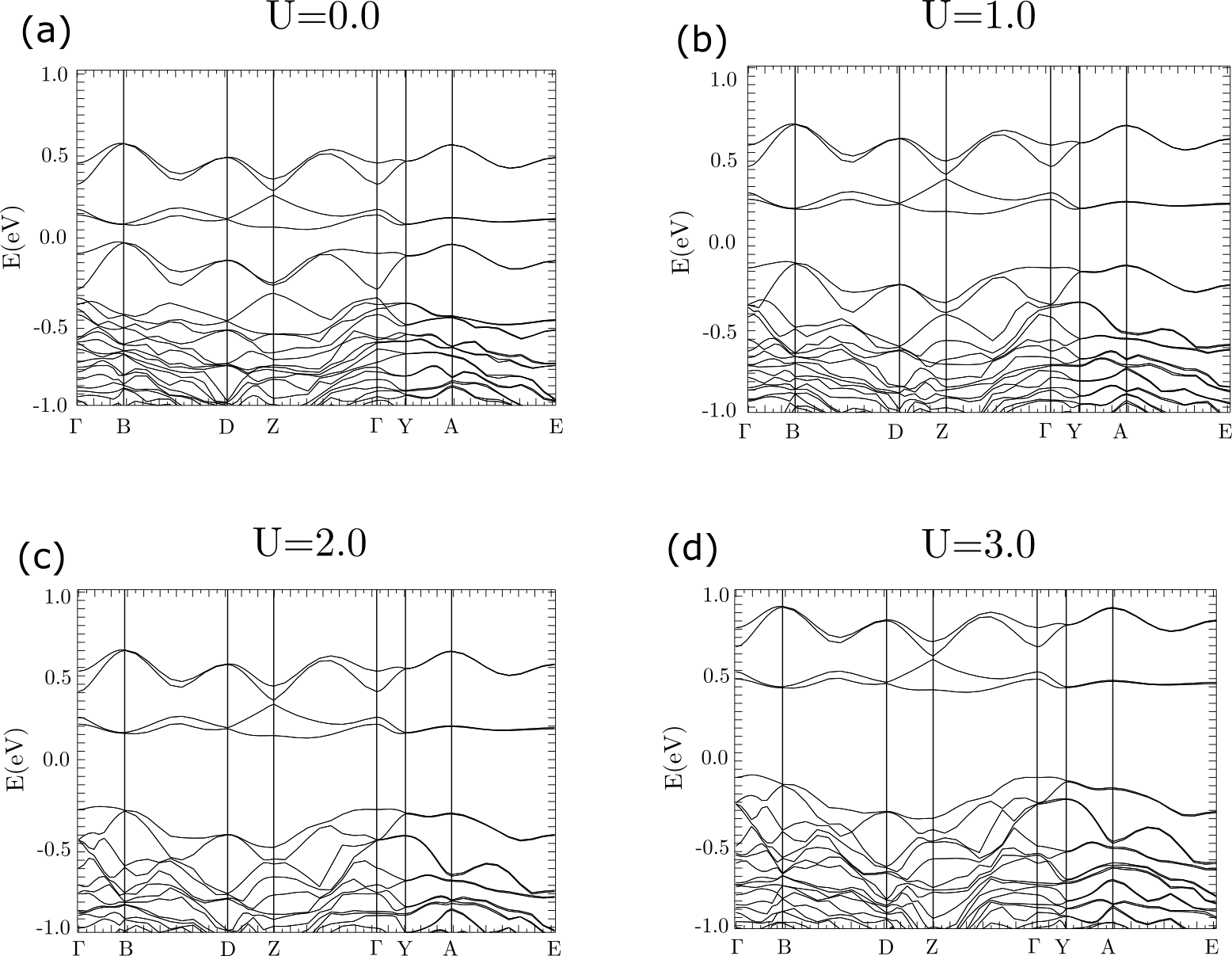}
	\caption{\label{phase}Computed {\it ab-inito} band structure of TlCuCl$_3$ with experimental coordinates in presence of spin-orbit coupling for different Hubbard $U$ for Cu 3d orbitals. }
\end{figure}

\section{SSH term}
The expression for SSH like term in our 3D model looks like
\begin{eqnarray}
H_{AA}&=C_{10} \cos((k_x+k_z)/2)+C_{11}\cos(k_y)+C_{12}\cos((k_x-k_z)/2)+C_{13}\cos(2*k_y)\\
&+C_{14}\cos(k_x)+C_{15}\cos(k_z)+C_{16}\cos(k_x+k_z)+C_{17}\cos(k_x-k_z)+C_{18}\nonumber\\
H_{AB}&=e^{ik_z/2}\left(C_0+C_1e^{-ik_z}+\left(C_2+C_3e^{-ik_z}\right)\left(C_4e^{i(k_x+k_y)/2}+C_5e^{-i(k_x+k_y)/2}\right)\right.\\
&\left.+\left(C_6+C_7e^{-ik_z}\right)\left(C_8e^{i(k_x-k_y)/2}+C_9e^{-i(k_x-k_y)/2}\right)\right)\nonumber
\end{eqnarray}
where C's are the fitting parameters. The above equation for inter-sublattice hopping can be rewritten in the form,
\begin{eqnarray}
H_{AB}=e^{i k_z/2}T_{\bf k\perp}(1+\frac{T'_{\bf k\perp}}{T_{\bf k\perp}}e^{ik_z})\nonumber
\end{eqnarray}

where
\begin{eqnarray}
T_{\bf k\perp}=C_0+C_2C_4e^{i(k_x+k_y)/2}+C_2C_5e^{-i(k_x+k_y)/2}+C_6C_8e^{i(k_x-k_y)/2}+C_6C_9e^{-i(k_x-k_y)/2}\nonumber\\
T'_{\bf k\perp}=C_1+C_3C_4e^{i(k_x+k_y)/2}+C_3C_5e^{-i(k_x+k_y)/2}+C_7C_8e^{i(k_x-k_y)/2}+C_7C_9e^{-i(k_x-k_y)/2}\nonumber\\
\end{eqnarray}
The fitting parameters are $C_{0-18}$=[0.156, 0.209, -0.076, 0.223, -0.135, -0.325, 1.466, -0.019, -0.030, 0.045, 0.002, -0.033, -0.003, 0.027, -0.002, 0.095, -0.105, -0.019] in eV.

\begin{figure}[h]
	\includegraphics[width=70mm,scale=3.5]{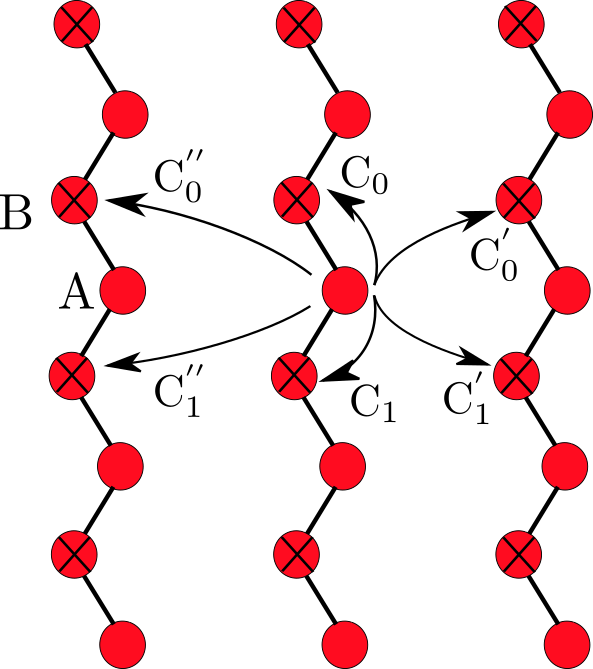}
	\caption{\label{phase}Figure showing effective hopping in (x+y,z) plane. Here $C_0^{'},C_0^{''},C_1^{'}$ and $C_1^{''}$ correspond to $C_2 C_4, C_2 C_5, C_3 C_4 $ and $C_3 C_6$ respectively. We have similar kind of hopping terme in (x-y,z) plane as well.}
\end{figure}

{
\section{Full form of SOC}
Although SOC is weak (as seen from the DFT calculations), however, it introduces chirality in the low energy spectrum across ${\bf k}^*$. Due to anisotropy between the $ab$-plane and the $c$-axis, we can conveniently split the SOC Hamiltonian into in-plane and out-of-plane, as 
\begin{eqnarray}
H_{\rm SO}&=&\sum_{i,j\in (A,B)}\sum_{{\bf k},ss'}\left[\psi^{\dag}_{i,s}({\bf k})\left({\bm \alpha}^{ij}_{\bf k}\times {\bm \sigma}_{ss'}\right)\psi_{j,s'}({\bf k}) + \psi^{\dag}_{i,s}({\bf k})\left({\bm \beta}^{ij}_{\bf k}\cdot\sigma^z_{ss'}\right)\psi_{j,s'}({\bf k})\right].
\label{Eq:HSO}
\end{eqnarray}
In the first term, the in-plane spin is locked to its transverse velocity matrix ${\bm \alpha}^{ij}_{\bf k}$, while the out-of-plane spin is locked to the longitudinal one ${\bm \beta}^{ij}_{\bf k}$. The components of the velicity operators are
\begin{eqnarray}
{\bm \alpha}^{ij}_{\bf k}& = &\alpha^{ij}_0\left(-\frac{\partial \xi^{ij}_{\bf k}}{\partial k_y},\frac{\partial \xi^{ij}_{\bf k}}{\partial  k_x},0\right), ~{\bm \beta}^{ij}_{\bf k}=\beta^{ij}_0\left(0,0,\frac{\partial \xi^{ij}_{\bf k}}{\partial k_z}\right). 
\label{Eq:SOC}
\end{eqnarray}
$\alpha^{ij}_0$, and $\beta^{ij}_0$ are the corresponding SOC strengths. Eq.~\ref{Eq:HSO} allows several SOC terms, however, fitting to DFT results indicate that $\beta^{ij}_0\rightarrow 0$, implying that the spins are aligned perpendicular to the SSH chain. $\alpha^{\rm AB}_0$ is the second nearest neighbor SOC term, and is negligibly small, while $\alpha^{\rm AA}_0=5$ meV.

}
\section{Chern-Simons-Ginzburg-Landau theory}
\subsection{Ginzburg-Landau theory}
Here we develop the Ginzburg Landau theory of our Hamiltonian around the antiferromagnetic order parameter.  We write the partition function for the total Hamiltonian $H+H_{\rm I}$ written in terms of the Dirac matrices in Eq.~(4) in the main text as
\begin{eqnarray}
Z=\int \mathcal{D}[\psi,\bar{\psi}]\exp\left[-\int_{0}^{\beta}d\tau\left(\bar{\psi}(\partial_\tau\mathbb{I}_{4\times 4}-H_{\bf k}) \psi - J\sum_{\langle{i,j}\rangle}S^z_i.S^z_j\right)\right],
\label{Eq:Z1}
\end{eqnarray}
where $\psi$ are 4 component Grassman variables $\psi=(\psi_{A\uparrow}, \psi_{B\uparrow},\psi_{A\downarrow},\psi_{B\downarrow})^T$ (equivalent to the Dirac spinor used in the main text), and $\bar{\psi}$ is the conjugate of $\psi$. $i$,$j$ denote `A', `B' sublattices. ${\bf S}_i$ are the corresponding spin operators. We orient the spin-quantization axis along $\sigma^z$, i.e., we only consider $S_i^z$ component. We define the antiferromagnetic (AF) field as $\phi = \left( S^z_{\rm A} - S^z_{\rm B}\right)/2$. Using the Hubbard Stratonovich transformation for $H_{\rm I}$ in terms of the FM fields in the last term of Eq.~\eqref{Eq:Z1}, we obtain
\begin{eqnarray}
\int \mathcal{D}[\psi,\bar{\psi}]\exp{\left(-J\sum_{\langle{i,j}\rangle}{\bf S}_i.{\bf S}_j\right)}&=&\int \mathcal{D}[\psi,\bar{\psi},\phi,\bar{\phi}]\exp\left[-J\phi(S^z_{{\rm A}}-S^z_{\rm B})-\frac{\phi^2}{4J}\right].
\label{Eq:Z1HI}
\end{eqnarray}

Now we express $S^z_i$ in terms of the Grassman variables as $S^z_i=(\bar{\psi}_{i\uparrow}\psi_{i\uparrow}-\bar{\psi}_{i\downarrow}\psi_{i\downarrow})/2$. In doing so, we can write the AF  term in terms of the Grassman spinor $\phi$ as $J\phi (S^z_{{\rm A}}-S^z_{\rm B}) = \bar{\psi}\left(J\phi\Gamma_5 \right)\psi$, where $\Gamma_5=\sigma_z\otimes\tau_z$, as defined in the main text. Substituting this identity in Eq.~\eqref{Eq:Z1HI}, and then inserting it back to Eq.~\eqref{Eq:Z1}, we get
\begin{eqnarray}
Z=\int \mathcal{D}[\psi,\bar{\psi},\phi,\bar{\phi}]\exp\left[-\int_{0}^{\beta}d\tau\bar{\psi}\left(\mathbf{G}_0^{-1}(\tau,{\bf k}) - \mathbf{M}(\phi)\right) \psi\right].
\label{Eq:Z2}
\end{eqnarray}
Here we have defined the non-interacting Green's function matrix $\mathbf{G}_0^{-1}({\bf k},\tau)=\partial_\tau\mathbb{I}_{4\times 4}-H_{\bf k}$, and the magnetization matrix as $\mathbf{M}(\phi)= J\phi\Gamma_5 $. Now we can go to the Matsubara frequency $i\omega_n$ domain and integrate out fermion variables ($\psi,\bar{\psi}$) to get the effective Lagrangian density as
\begin{eqnarray}
\mathcal{L}={\rm Log} \left[{\rm Det} \left(\sum_{{i\omega_n},{\bf k}} \mathbf{G}_0^{-1}(i\omega_n,{\bf k})-\mathbf{M}(\phi)\right)\right]-\frac{\phi^2 }{4J}.
\end{eqnarray}

Under the saddle point approximation around the AFM, using the identity $\text{Log[Det}[..]]=\text{Tr[Log}[..]]$ and $\text{Log}[x]=-\sum_{n=1}^{\infty}(-1)^nx^n/n$, we get the GL Lagrangian potential
\begin{eqnarray}
\mathcal{L}_{\rm GL}= -\alpha|\phi|^2-\beta|\phi|^4+\mathcal{O}(|\phi|^6),
\label{LGL}
\end{eqnarray}
where 
\begin{eqnarray}
\alpha &=&-\frac{1}{4J}+{\rm Tr}\sum_{k,k'} {\bf G}_0(k)\Gamma_5 {\bf G}_0(k')\Gamma_5,\\
\beta &=& {\rm Tr}\sum_{k,k',k'',k'''}{\bf G}_0(k)\Gamma_5 {\bf G}_0(k')\Gamma_5 {\bf G}_0(k'')\Gamma_5 {\bf G}_0(k''')\Gamma_5,
\end{eqnarray}
where we define $k=({\bf k},i\omega_n)$. Exact computation of $\alpha$, and $\beta$ variables are difficult, but we we can already grasp the essence that $\beta>0$, and $\alpha\rightarrow 0$ when the particle-hole bubble compensates the interaction terms. These results are typical for the GL theory.

\subsection{The Chern-Simons term}
Chern-Simons term arises in the presence of electromagnetic (EM) fields. We assume the probe electromagnetic fields as $(A_0, {\bf A})$. In addition to probe fields, there may arise intrinsic `statistical' gauge fields $(a_0, {\bf a})$ due to fluctuations of the bosonic fields $\phi$. This can be seen easily. The statistical gauge field arises due to fluctuations of the order parameter, so we can write $a_0\propto \partial_t (\delta \phi\delta \phi)$, and ${\bf a}\propto \nabla (\delta \phi\delta \phi)$, where $\delta \phi$ is the fluctuation of the AFM field around its saddle point $\phi_0$. Such intrinsic gauge clearly arises from the $|\phi|^4$ term in the GL potential in Eq.~\eqref{LGL}, and persists above the AFM critical point. We are not particularly interested in the details of the origin of the intrinsic guage field, except it conveys an important message that such due to spin-fluctuations in space-time dimensions, there can be CS term even in the absence of any external EM field. Readers interested in the details of the origin of such statistical gauge field can refer to Refs.~\cite{CSGL_Zhang,axionJPSJ} and references therein

Thanks to the linear combination form of the intrinsic and probe gauge fields in the Lagrangian, we can combine their effects in a total gauge field as $\mathcal{A}_0=a_0+A_0$, and $\bm{\mathcal{A}}={\bf a}+{\bf A}$. Due to the total EM field, we have a typical Maxwell term ($\mathcal{L}_{\rm MW}$), and the Chern-Simons term $\mathcal{L}_{\rm CS}$ as defined in 3+1 dimensions as\cite{zhang_prb,JMoore,CSGL_rest}
\begin{eqnarray}
\mathcal{L}_{\rm MW}&=&-\frac{1}{4}\mathcal{F}_{\mu\nu}\mathcal{F}^{\mu\nu}-\mathcal{A}_{\mu}\mathcal{J}^{\mu},\\
\mathcal{L}_{\rm CS}&=& \theta\frac{\hbar}{\Phi_0^2}\epsilon^{\mu\nu\sigma\tau}\partial_{\mu}\mathcal{A}_{\nu}\partial_{\sigma}\mathcal{A}_{\tau}-\mathcal{A}_{\mu}\mathcal{J}^{\mu}.
\label{CS}
\end{eqnarray}
where the Einsteins summation convention is implied. $\mathcal{F}_{\mu\nu}=\partial_{\mu}\mathcal{A}_{\nu}-\partial_{\nu}\mathcal{A}_{\mu}$, current density $\mathcal{J}^{\mu}$ is included by conservation principles and can be eliminated for the Lagrangian minimization problem of our interest. $\theta$ is the axion angle which is related to the momentum-space non-Abelian Berry connection $\mathscr{A}^{st}_{\mu}=-i\langle u^{s}_{\bf k}|\partial_{k_{\mu}}u^{t}_{\bf k}\rangle$, where $|u^s_{\bf k}\rangle$ is the $s^{\rm th}$-eigenstate of the mean-field Hamiltonian,\cite{zhang_nat} as
\begin{eqnarray}
\theta=\frac{1}{4\pi}\int_{\rm BZ}d^3k \epsilon^{\mu\nu\sigma}{\rm Tr}\left[{\mathscr{A}}_{\mu}\partial_{\nu}{\mathscr{A}}_{\sigma}+i\frac{2}{3}{\mathscr{A}}_{\mu}{\mathscr{ A}}_{\mu}{\mathscr{A}}_{\mu}\right].
\end{eqnarray}
By evaluating the eigenvectors of our Hamiltonian in the main text, we can obtain an algebric, gauge independent form of the axion angle can be deduced to be:
\begin{eqnarray}
\theta=\int_{\rm BZ}\frac{d^3k}{4\pi} \frac{2|d|+d_4}{(|d|+d_4)^2|d|^3}\epsilon^{ijkl}d_i\partial_{k_x} d_j\partial_{k_y} d_k\partial_{k_z} d_l.
\label{ax_2}
\end{eqnarray}
where $|d|^2=\sum_{i=1}^5|d_i|^2$, and $d_5=J\phi$, and $i,j,k,l$ runs from 1,2,4,5. The above integral evaluates the solid angle enclosed in the $d$-space as one encircles the entire 3D Brillouin zone in the $k$-space. Reminiscence to the topological phase transition in a single SSH chain, here also we can show that $\theta$ acquires finite value where the zeros of $d_3({\bf k})$ term lies inside the solid angle, giving the condition that $T_{{\bf k}_{\perp}}\le T'_{{\bf k}_{\perp}}$, for ${\bf k}\in {\rm BZ}$. Having a Dirac cone in the SOC band structure, we ensure that such a condition is automatically satisfied in the non-interacting phase ($d_5=0$). Axion angle can be calculated numerically. We are interested in studying its behavior as a function AFM field $\phi$, which yields an exponential function $\pi e^{-\lambda|\phi|}$, where $\lambda$ is a fitting parameter. For both signs of $\phi$, $\theta$ decreases from $\pi$ at $\phi\rightarrow0$. Absorbing the remaining factors in the CS term into  $\gamma=\frac{\pi\hbar}{\Phi_0^2}{\bf E}\cdot{\bf B}$, we obtain $\mathcal{L}_{\rm CS}=\gamma e^{-\lambda|\phi|}$. $\gamma>0$ ($\gamma<0$) if ${\bf E}$ and ${\bf B}$ are parallel (antiparallel) to each other.    

\subsection{Chern-Simons-Ginzburg-Landau theory} 
The Kinetic energy term due to the AFM field is 
\begin{eqnarray}
\mathcal{L}_{\rm KE} &=& i\phi^*\mathcal{D}_0\phi + \frac{1}{2m}\phi^*\bm{\mathcal{D}}^2\phi.
\label{Eq:KE}
\end{eqnarray}
Here the covariant derivative operators are $\mathcal{D}_0=\partial_t +ie\mathcal{A}_0$, and $\bm{\mathcal{D}}=i{\bm \nabla}+e\bm{\mathcal{A}}$. Therefore the total Lagrangian density becomes\cite{CSGL_rest,CSGL_Zhang} $\mathcal{L}_{\rm total}=\mathcal{L}_{\rm KE} + \mathcal{L}_{\rm MW} + \mathcal{L}_{\rm GL} + \mathcal{L}_{\rm CS}+ \mathcal{L}_{\rm AN}$. Here $\mathcal{L}_{\rm AN}$ represents the contribution from anyons arising from the fluctuation of the order parameters. The Maxwell term does not involve the order parameter or axion term, and thus also can be neglected. Neglecting space-time dependence of the order parameter, we otain the effect CSGL term in terms of the AFM field $\phi$ as
{
\begin{eqnarray}
\mathcal{L}_{\rm CSGL}&=&-\alpha|\phi|^2 - \beta|\phi|^4 - \gamma e^{-\lambda|\phi|}+\gamma.
\label{CSGL}
\end{eqnarray}
}

\section{Free energy minima and Transition temperature}
The Free energy is minimum where the Lagrangian in Eq.~\eqref{CSGL} is maximum. Solving for $\frac{\partial F[\phi]}{\partial \phi}|_{\phi_0}=0$, we get $2\left(\alpha+2\beta|\phi_0|^2\right)|\phi_0| = \gamma\lambda e^{-\lambda|\phi_0|}$. This equation cannot be solved analytically, but for small $\lambda$, we can expand the exponential up to third power in $\phi$ to get 
\begin{eqnarray}
2\alpha|\phi_0|+4\beta|\phi_0|^3-\lambda \gamma (1-\lambda|\phi_0|+\frac{\lambda^2|\phi_0|^2}{2!}-\frac{\lambda^3|\phi_0|^3}{3!})=0.
\label{saddlept2}
\end{eqnarray}
It turns out that any arbitrarily small value of $\gamma$, this leads to a minima in free energy away from $\phi=0$ but very close. At high temperature, it goes arbitrarily close to 0. 

The axion term leads to a correction to the N\'eel temperature $T_{\rm N}$. If we assume $T_{\rm N,0}$ is the N\'eel temperature without the axion term, then for a second order phase transition, we can write { $\alpha=\alpha_0(1-T/T_{\rm N,0})$}, where $\alpha_0>0$ is a constant. This coefficient is modified to  
$\alpha'=\alpha+\gamma\lambda^2/2$. $\alpha'=0$ gives the AFM transition. Therefore, the phase transition condition becomes $\alpha_0\left(1-\frac{T_{\rm N}}{T_{\rm N,0}}\right) + \gamma\lambda^2/2 = 0$, which gives $T_{\rm N}=T_{\rm N,0}\left(1+\frac{\gamma\lambda^2}{2 \alpha_0}\right)$.

\section{ Higgs Mass and Lifetime}
Using CSGL Lagrangian, we can calculate the mass of the Higgs mode. This can be done by substituting an amplitude fluctuation term $\delta\phi$ in the Lagrangian, and calculating the coefficient of the quadratic term in fluctuations. (We do not worry about the Goldstone modes here since they are guaged awayout from the Lagrangian.) Replacing $|\phi|\to |\phi_0|+\delta \phi$ in Eq.~\eqref{CSGL} (assuming $\phi$ is positive), evaluating $\partial^2 \mathcal{L}_{\rm CSGL}[|\phi_0|+\delta \phi]/\partial^2 \delta \phi$ at $\delta \phi\to 0$, we get the Higgs mass as
{
\begin{eqnarray}
M=\frac{\partial^2 \mathcal{L}_{\rm CSGL}}{\partial^2 \delta \phi}\Big|_{\delta\phi=0}&=&-2\alpha-12\beta|\phi_0|^2+\gamma\lambda^2e^{-\lambda |\phi_0|},\nonumber\\
&=&-2\alpha  -2\lambda\alpha|\phi_0|-12\beta|\phi_0|^2 -4\lambda\beta|\phi_0|^3.
\end{eqnarray}
}
In the last equation, we have substituted the condition for the saddle point, given above Eq.~\eqref{saddlept2}.

Unlike Higgs mass, its lifetime cannot be estimated exactly. One source of lifetime to the Higgs boson is the interaction term, i.e. quartic term in the Lagrangian. In this spirit, the leading term in the inverse lifetime ($\tau$) of the Higgs mass is proportional to the coefficient of the $\delta\phi^4$, which can be calculated by setting $\delta\phi=0$ in $\partial^4 \mathcal{L}_{\rm CSGL}[|\phi_0|+\delta \phi]/\partial \delta \phi^4$ which leads to 
{
\begin{eqnarray}
\frac{1}{\tau}&\propto& 24 \beta|\phi_0|+\lambda^4\gamma e^{-\lambda|\phi_0|},\nonumber\\
&=& 24\beta+2\lambda^3\phi_0\left(\alpha+4\beta\phi_0^3\right).
\end{eqnarray}
}
Here one should take the absolute value of the right-hand side.